\documentclass[aps,twocolumn,prl,amsmath,amssymb,10pt,aps]{revtex4-1} 
\usepackage{graphicx}
\usepackage{latexsym}
\usepackage{amsmath}
\usepackage{amsthm}
\usepackage{amsbsy}
\usepackage{amssymb}
\usepackage{epstopdf} 
\usepackage{enumerate}
\usepackage{setspace} 
\usepackage{dcolumn}
\usepackage{bm}
\usepackage{setspace} 
\usepackage{slashed}
\usepackage{color}
\usepackage{youngtab}
\usepackage[colorlinks=true]{hyperref}
\usepackage{float}
\usepackage{graphicx}
\usepackage{multirow}
\usepackage{array}
\newcolumntype{L}[1]{>{\raggedright\arraybackslash}p{#1}}
\newcolumntype{C}[1]{>{\centering\arraybackslash}p{#1}}
\newcolumntype{R}[1]{>{\raggedleft\arraybackslash}p{#1}}

\usepackage{color}
\usepackage{makecell}

\begin{document}
  \title{  Exotic Thermal Transitions with Spontaneous Symmetry Breaking
  }
  \author{ Hanbit Oh }
 \author{Eun-Gook Moon }
\thanks{egmoon@kaist.ac.kr}
\affiliation{Department of Physics, Korea Advanced Institute of Science and Technology, Daejeon 305-701, Korea}

\date{\today}

\begin{abstract}  
We show that exotic spontaneous symmetry breaking appears in thermal topological phases by perturbing the exact solutions of {\it quantum} rotor models coupled to the three-dimensional toric code. 
The exotic Ising and XY transitions are shown to be in the same universality class in drastic contrast to the conventional Wilson-Fisher classes without topological orders. 
Our results indicate that topological orders must be included to pin down universality classes of thermal transitions in addition to order parameter symmetry and spatial dimension. 
We evaluate all the critical exponents and find that the exotic universality class is more stable under the couplings to acoustic phonons and disorder.
Applying our results to experiments, we provide a plausible scenario in puzzlings of strongly correlated systems, including the absence of specific heat anomaly in doped RbFe$_2$As$_2$.  
 \end{abstract}
 
\maketitle 

{\it Introduction:} 
The phenomenological Landau theory is one of the most successful theories in physics that explain phase transitions by introducing the concept of the order parameter \cite{GL1,GL4}.
Including fluctuations of order parameters, the universality classes of continuous phase transitions are discovered with the development of renormalization group analysis, so-called Wilson-Fisher (WF) classes \cite{RG2,RG3}.
The conventional wisdom that the WF classes are solely determined by order parameter symmetry and spatial dimension is established.

Striking and perplexing phenomena are reported even in recent experiments of thermal transitions of high-temperature superconducting materials.
The Ising nematicity in layered cuprates, doped YBa$_2$Cu$_3$O$_7$ and HgBa$_2$CuO$_{4}$, shows a peculiar super-linear onset \cite{super1,super2,super3,super4,super5,super6}.
In doped RbFe$_2$As$_2$, the diagonal Ising nematicity without anomaly in heat capacity is observed \cite{Rb1,Rb2,Rb3,Rb4}.
The conventional WF classes are inapplicable to these experiments, calling for a new theoretical framework.  
Possibilities of exotic transitions have been suggested by using the ideas of topology, fractionalization, and deconfinement \cite{KT1,KT2,QSL1,QSL2,DC1,DC2,Zhenbi,Moon2,Oh1}.
 
In this work, we demonstrate the existence of exotic thermal transitions by investigating spontaneous symmetry breaking transitions in thermal phases with topological orders. 
Instead of using conventional thermal gauge theories \cite{kogut, higgs_Fradkin, heisenberg1,heisenberg2}, 
we analyze the quantum rotors coupled to qubits of the toric code in three spatial dimensions (3d) \cite{TC2d_1,TC3d_1,TC3d_2} to explore exotic thermal transitions. 
Striking characteristics of the exotic thermal transitions are uncovered.  
The Ising and XY transitions are in the same universality class in drastic contrast to the WF classes. 
All critical exponents are evaluated, and their differences from the ones of the WF classes are emphasized. 
 
Our approach has the following advantages. 
First, there is no imposed gauge invariance in our models and thus no subtlety associated with the origin of the gauge invariance in thermal phases.
In fact, we discuss the effects of gauge non-invariant interactions and show their irrelevance to the existence of the exotic universality classes.
Second, we utilize the exact solutions of the models and employ controlled perturbative analysis. 
The clear-cut conclusions of the existence and properties of exotic thermal transitions are obtained.  
Couplings to other physical degrees of freedom, such as acoustic phonons or Fermi-surfaces, are determined unambiguously, and the stability of the thermal transitions are studied.
Third, our quantum models naturally provide the relations between the thermal phase transitions and corresponding quantum phase transitions, providing a bird's-eye view of quantum and thermal transitions.  \\

{\it Stability of the WF classes :} 
We revisit the WF universality classes of thermal phase transitions by considering {\it quantum} rotor models. 
This is not only for setting up our notations but also for considering the stability of the WF classes under fermions with Fermi-surfaces and acoustic phonons.  
To be specific, let us consider the $O(2)$ quantum rotor model on a cubic lattice with the conventional hat notation for quantum operators, 
\begin{eqnarray}
  \hat{H}_{R} &=&\sum_j \frac{\textcolor{black}{V_{\theta}}}{2} \hat{n}_j^2  -  t_{\theta} \sum_{\langle i, j \rangle}  \cos( \hat{ \theta}_i - \hat{ \theta}_j), \nonumber 
  \end{eqnarray}
  where an angle operator, $\hat{\theta}_j$, and its conjugate number operator, $\hat{n}_j$, with the commutation relation, $[e^{i \hat{\theta}_j}, \hat{n}_j]=e^{i \hat{\theta}_j}$, are introduced.
  The Hilbert space is a tensor product of local Hilbert spaces, 
  \begin{eqnarray}
  \mathcal{H}_{R} = \prod_j \otimes \mathcal{H}_j, \quad \mathcal{H}_j = \{ | n_j \rangle \,\, |  \,\, n_j \in \mathbb{Z} \}, \nonumber
  \end{eqnarray}
  with $  \hat{n}_j | n_j \rangle = n_j  | n_j \rangle $. 
  The integer condition of $n_j$ is associated with the compactification, $| \theta_j \rangle = | \theta_j +2\pi \rangle$. 
  The model enjoys a $U(1)$ symmetry whose action is $\hat{U}(\alpha) = \prod_j e^{i \alpha \hat{n}_j}$ with a real value $\alpha$. 
  The symmetry is spontaneously broken by varying with $\textcolor{black}{V_{\theta}}/t_{\theta}$.
  The symmetric phase is adiabatically connected to the ground state at $t_{\theta}=0$, $|{\rm sym} \rangle = \prod_j | n_j =0\rangle$, and the symmetry broken phase is adiabatically connected to the ground state at $\textcolor{black}{V_{\theta}}=0$, 
  \begin{eqnarray}
   | \theta_0  \rangle = \prod_j  | \theta_j=\theta_0 \rangle, \quad 
\langle  \theta_0   | e^{i \hat{\theta}_j} | \theta_0  \rangle/\langle  \theta_0   | \theta_0  \rangle =e^{i \theta_0} , \nonumber
  \end{eqnarray} 
  for $\theta_0 \in(0, 2\pi)$.
  Thermal fluctuations at temperature $T$ drives a thermal phase transition of $U(1)$ symmetry breaking. 
  One direct way to construct the corresponding Landau functional is to introduce a complex order parameter, $\psi_j = e^{i \theta_j}$, and perform a coarse-graining, which gives
  \begin{eqnarray}
&&\mathcal{Z}_{R} ={\rm Tr}_{\theta} (e^{- \hat{H}_{R}/T} ) \quad \rightarrow \quad \mathcal{Z}_{R} \simeq \int D \psi \, \, e^{- \mathcal{F}_{R}(\psi)  }, \nonumber \\
&&   \mathcal{F}_{R}[\psi]= \int_x \Big( |\nabla \psi(x)|^2 + r |\psi(x) |^2 + \frac{\lambda}{4}|\psi(x)|^4 \Big). \nonumber 
\end{eqnarray}  
Here, $\psi(x)$ is coarse-grained field and the integration is over a 3d space. Hereafter, we omit obvious space dependence of order parameters. The coupling constants ($r,\lambda$) are functions of ($\textcolor{black}{V_{\theta}}, t_{\theta}, T$) in addition to a lattice spacing. 
The functional describes the WF-XY universality class in 3d, and the $U(1)$ symmetry breaking transition is equivalently understood by a proliferation of the primary topological defects, $2\pi$ vortex-lines.

It is well-known that the WF-XY class is unstable under the Ising potential, $V_{I}=- u \sum_j \cos( \theta_j)^2$, and the presence of a non-zero $u$ breaks the $U(1)$ symmetry down to the Ising one whose order parameter becomes a real variable, $\phi_j = \cos(\theta_j)$. The Ising potential has the form, $V_{I}=- u \sum_j \phi_j^2$, and the Landau-functional of the Ising order parameter becomes 
\begin{eqnarray}
 \mathcal{F}_{I}[\phi]& =& \int_x  \Big(  \frac{1}{2} (\nabla \phi)^2 + \frac{r}{2} (\phi)^2  + \frac{\lambda}{4} ( \phi)^4 \Big). \nonumber
\end{eqnarray}

Next, let us investigate how the theory of the order parameter is affected by fermions with Fermi-surfaces. 
The most drastic effects come with the presence of Fermi-surface, so we consider a model Hamiltonian, $H= H_{I} + H_{f} + H_{Y}$ and
\begin{eqnarray}
 H_{f}=  -  t_f \sum_{\langle i, j \rangle}  c_i^{\dagger} c_j -\mu \sum_j c^{\dagger}_j c_j , \  H_{Y} =  -g \sum_{j} \phi_j c_j^{\dagger}\hat{M} c_j ,\nonumber 
 \end{eqnarray}
 where $\hat{M}$ is a vertex operator of the Yukawa coupling. 
After integrating out the electrons, the partion function is written schematically, 
\begin{eqnarray}
\mathcal{Z} = {\rm Tr}_{\phi}\Big[e^{-\frac{H_{I}}{T}} {\rm det} \big[i \omega_n-\epsilon_k(t_f,\mu) +g \phi_j \hat{M} \big] \Big]. \nonumber
\end{eqnarray}
The fermion Matsubara frequency is  $\omega_n = (2n+1)\pi T $, and no singularity appears in the expansion at non-zero temperatures.
For example, the interaction coefficient is modified by electrons, $u \rightarrow u+ \Pi(\mu,t_f)$.
Note that the absence of singularities from electrons indicates that singularities in the free energy are only associated with phase transitions (See Supplemental Material (\textcolor{cyan}{SM})\cite{SM}). 
Thus, the same criticality theory with modified interaction terms describes phase transitions. The Ising class is stable under the coupling with fermions.

The stability of the WF classes under acoustic phonons has been well understood in the literature \cite{LP1,LP2,LP6,jorg}. 
Introducing a phenomenological coupling through a strain tensor, the Larkin-Pikin condition is suggested by the lowest order renormalization group analysis, showing that the Ising (XY) universality becomes unstable (stable) in 3d, respectively. 
The Ising universality would become either a first-order transition or a different universality class such as the mean-field class. \\

{\it The Model :} 
We couple quantum rotors to qubits $\hat{\sigma}_l^{x,y,z}$ at links of a 3d cubic lattice with a periodic boundary condition. 
The total Hilbert space becomes the tensor product of the ones of quantum rotors and qubits,  
\begin{eqnarray}
\mathcal{H}_{{\rm tot}} = \mathcal{H}_R \otimes \mathcal{H}_{Q}, \quad \mathcal{H}_{Q}= \prod_{l} \otimes \{  | \sigma_l^z =\pm1 \rangle\}. \nonumber
\end{eqnarray}
The model Hamiltonian is
\begin{eqnarray}
\hat{H}_X =  -  J \sum_{\langle i, j \rangle} \hat{\sigma}_{ij}^z \cos( \frac{\hat{\theta}_i}{2} -\frac{\hat{\theta}_j}{2}) -\sum_j  e^{i 2\pi \hat{n}_j} \hat{\mathcal{A}}_j - \sum_{p^{*}} \hat{\mathcal{B}}_{p^*} , \ \ \ \ 
\end{eqnarray}
where the star and plaquette operators, $ \hat{\mathcal{A}}_j = ( \prod_{l \in j} \hat{\sigma}_l^x)$, $\hat{\mathcal{B}}_{p^*} = (\prod_{l \in p^*} \hat{\sigma}_l^z)$ of the 3d toric code are introduced as shown in Fig.\ref{F1}\textcolor{red}{(a)}.
The index $l$ and $j$ are for a link and a site, respectively.
We note that the original 3d toric code has a thermal phase transition at $\Lambda^0_c=1.313$, which describes a deconfinement-confinement transition \cite{TC3d_1}.  

Few remarks are as follows. 
The $U(1)$ symmetry of $\hat{H}_X$ is the same as the one in $\hat{H}_R$.
The factor $e^{i 2\pi \hat{n}_j} $ associated with the star operator is an identity operator due to the integer condition of $n_j$, still, its presence is useful to check symmetry apparently.  
Also, the model is exactly solvable because all the terms of $\hat{H}_{X}$ commute with each other, which is related to the local $\mathbb{Z}_2$ transformation generated by $e^{i 2\pi \hat{n}_j} \hat{\mathcal{A}}_j$.
The effects of other interactions which break the exact solvability and local $\mathbb{Z}_2$ transformation are discussed below.

\begin{center}
\begin{figure}[t]
\includegraphics[scale=1.4]{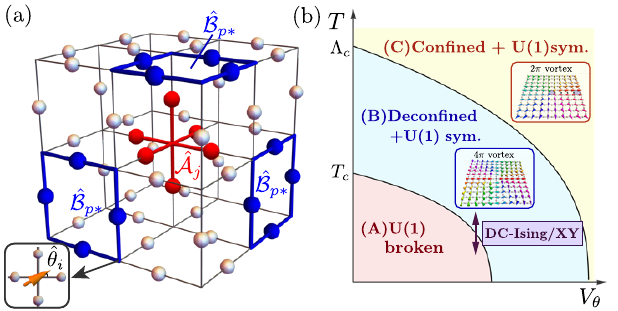}\quad\ 
\vspace{-8pt}
\caption{\textbf{(a) Interaction terms of $\hat{H}_X$.}
One of the star operators $\hat{\mathcal{A}}_{j}$ and three of plaquette operators $\hat{\mathcal{B}}_{p*}$ of the 3d toric code are denoted with red and blue links, respectively. 
\textbf{(b) Schematic phase diagram.}
Two transitions $(\Lambda_{c}, T_{c})$ correspond to a symmetry breaking and deconfinement. 
The DC-Ising/XY is different from the conventional WF-Ising/XY, whose critical exponents are in Table \ref{T2}. 
 }
 \label{F1}
 \vspace{-10pt}
\end{figure}
\end{center}

One ground state with a quantum number $\theta_0$ is, 
\begin{eqnarray}
| {\tilde \theta}_0 \rangle =\hat{\mathcal{P}}_{G} \big( | \theta_0 \rangle \otimes  \prod_l | \sigma_{l}^z=1\rangle \big), \  \hat{\mathcal{P}}_{G} \equiv  \prod_j \big(\frac{1+ e^{i 2\pi \hat{n}_j} \hat{\mathcal{A}}_j}{2} \big).\nonumber
\end{eqnarray}
Here, $\hat{\mathcal{P}}_G$ is a projection operator onto the states with $\mathcal{A}_{j}=1$ for all $j$.
We stress that the order parameter is the expectation value of an operator $e^{i \hat{\theta}_j}$, not $e^{i \hat{ \theta}_j/2}$, as manifested by 
\begin{eqnarray}
\frac{\langle \tilde{ \theta}_0 | e^{i \hat{\theta}_j} | \tilde{ \theta}_0 \rangle}{\langle \tilde{ \theta}_0 | \tilde{ \theta}_0 \rangle} =e^{i \theta_0}, \quad\frac{ \langle  \tilde{ \theta}_0 | e^{i \hat{\theta}_j/2} | \tilde{ \theta}_0 \rangle}{\langle \tilde{ \theta}_0 | \tilde{ \theta}_0 \rangle}  =0. \label{order}
\end{eqnarray}
The ground state with a quantum number $\theta_0$ is unique without any degeneracy. 
For example, we consider the spin-flip operator along the crystal plane $\mathcal{L}_{xy}$ of the dual lattice perpendicular to z-direction, $\hat{\mathcal{V}}_{xy}= \prod_{l \in \mathcal{L}_{xy}} \hat{\sigma}_l^x$ (See Fig.\ref{SF2}).
While it connects homologically different ground-states in the pure 3d toric code, the operator action in our model gives the energy,
\begin{eqnarray}
\frac{\langle {\tilde \theta}_0 |\hat{\mathcal{V}}_{xy} \hat{H}_X \hat{\mathcal{V}}_{xy} | {\tilde \theta}_0 \rangle }{ \langle \tilde{ \theta}_0 | \tilde{ \theta}_0 \rangle}= E_G + 2J L_x^2, \nonumber 
\end{eqnarray}
lifting the degeneracy completely. 
Here, $E_G$ is for the ground state energy without any spin-flip operator.

The wave functions and excitation energies are summarized in Table \ref{T1},\ref{ST1}, and a few remarks are as follows. 
First, the primary topological defect is a $4\pi$ vortex-line, not a $2\pi$ vortex-line \cite{gauge_Subir}. 
Note that the bound state of a $2\pi$ vortex-line and a flux-line has lower energy, but it is still qualitatively bigger than the energy of a $4\pi$ vortex-line. 
Second, the energy hierarchy manifests. 
Only phase fluctuations and $4\pi$ vortex live at low energy Hilbert space.  
Hereafter, we consider the case with $J, T \ll \Lambda_c^0$, which allows us to focus on the zero flux sector.

{\it Exotic Transitions associated with Topological Orders:} 
In the zero flux sector, the effective Hamiltonian is 
\begin{eqnarray}
\hat{H}_{{\rm eff}} =\hat{ \mathcal{P}}_{0} \hat{H}_X\hat{\mathcal{P}}_{0} \rightarrow -  J \sum_{\langle i, j \rangle}  \cos( \frac{\hat{\theta}_i}{2} -\frac{\hat{\theta}_j}{2}), \nonumber
\end{eqnarray}
where the projection operator onto the zero flux Hilbert space, $\hat{\mathcal{P}}_{0}$, is introduced. 
The right hand side of the arrow is obtained with the configuration of the zero flux, $\sigma_{l}^z=1$.
Remark that the range of the angle variables is $\theta_j \in (0,2\pi)$, originated from $\mathcal{H}_R$, not the doubled one $(0, 4\pi)$.

The form of the effective Hamiltonian indicates that the $2\pi$ vortex-lines are also removed from the zero flux Hilbert space. 
Thus, the zero-flux Hilbert space can be written as
\begin{eqnarray}
\hat{\mathcal{P}}_{0} [ {\mathcal{H}}_{{\rm tot}}] =\mathcal{H}_{0\pi_{v}} \otimes \mathcal{H}_{4\pi_{ v}}\otimes \mathcal{H}_{8\pi_{ v}} \otimes \cdots, \nonumber
\end{eqnarray}
where $\mathcal{H}_{4n\pi_{ v}}$ is for the Hilbert space with $4n\pi  $ vortex-lines. 
Then, the corresponding partition function is, 
\begin{eqnarray}
\mathcal{Z}_{{\rm eff}} \equiv {\rm Tr}_{\theta}\big[e^{-\hat{H}_{{\rm eff}}/T}\big] = \mathcal{Z}_{0\pi_{ v}} \mathcal{Z}_{4\pi_{ v}} \cdots, \nonumber
\end{eqnarray}
where the subscripts are to specify the topological defects. 
Then, the remaining step is standard. 
A topological phase transition associated with $4\pi$ vortex-lines appears in $\mathcal{Z}_{4\pi_{ v}}$, whose critical temperature is estimated by comparing energy and entropy of the topological defect, and we find the critical temperature, $T_c \sim J$.

\begin{table}[t]
\renewcommand{\arraystretch}{1.38}
\begin{tabular}{C{0.43\linewidth}C{0.52\linewidth}}
\hline 
\hline
Excitation & Energy cost  \\ \hline
 A pair of fluxes    & $8+2J$   \\ 
$2\pi\;$vortex  &  
$(\frac{\pi}{4}\log
 \left( \frac{L_x}{a}\right)+\frac{L_x}{a})J\frac{L_x}{a}$   \\
 $2\pi\;$vortex + fluxes   &    $\left(\frac{\pi}{4} \log
 \left( \frac{L_x}{a}\right)J+4\right) \frac{L_x}{a}$  \\ 
 $4\pi\;$vortex   &    $(\pi \log
 \left( \frac{L_x}{a}\right))J\frac{L_x}{a}$\\
Phase fluctuation & $J(1-\cos(\frac{\alpha_j}{2}))$    
\\ \hline \hline
\end{tabular}
\caption{\textbf{Excitations and their excitation energies of $\hat{H}_X$. }
 The vortex and flux configurations of the rotor and qubit on the lattice are illustrated in Fig.\;\ref{SF3}.
} \label{T1}
\end{table} 
 
We stress that the trace of $\mathcal{Z}_{{\rm eff}}$ is over $\{ | \tilde{\theta_0} \rangle \}$ not over the conventional states, $\{ |\theta_0 \rangle \}$, and there is no ambiguity of the thermal average condition, $ \langle e^{i \hat{\theta}_j/2} \rangle_T =0$. 
It is also important to note that the partition function becomes asymptotically exact in the limit $J, T \ll \Lambda_{c}^0$, and the phase transition appears varying with $J/T$.   

The coarse-grained Landau functional is obtained by introducing the complex variable,  $\psi_j = e^{i \theta_j/2}$, 
\begin{eqnarray}
\mathcal{F}_{{\rm DC}}[\psi]= \int_x \Big( |\nabla \psi|^2 + r |\psi|^2 + \frac{\lambda}{4}|\psi|^4 \Big). \nonumber 
\end{eqnarray} 
The universality class of $\mathcal{F}_{{\rm DC}}[\psi]$ is different from the one of $\mathcal{F}_{R}[\psi]$ even though they have the same form because the variable,  $\psi_j $, is not an order parameter. Instead,  a secondary operator of the variable, $\psi_j^2$, is an order parameter as shown in Eq. \eqref{order}. 
Then, we employ the thorough investigations by Vicari and collaborators to understand $\mathcal{F}_{{\rm DC}}[\psi]$ around $T_c$ \cite{Vicari}. 
The correlation length critical exponent with $\xi \sim (T_c -T)^{-\nu}$ is estimated as $\nu=0.67$, and the order parameter scaling dimension with $ \langle (\psi)^2 +(\psi^{\dagger})^2 \rangle_T \sim (T_c -T)^{\beta}$ is estimated by the scaling dimension of the secondary operator, $\beta=0.83$. 
With the two independent critical exponents, all other critical exponents are obtained by the scaling relations, which are summarized in Table \ref{T2}. 

Let us consider the Ising potential at $T_c$. The potential in terms of $\psi_j$ is 
\begin{eqnarray}
V_I = -u \sum_j \cos(\theta_j)^2 = - u \sum_j (\psi_j^2 + (\psi_j^{\dagger})^2)^2, \nonumber
\end{eqnarray} 
and add $\int_x (\psi^2 +(\psi^{\dagger})^2)^2$ to $\mathcal{F}_{{\rm DC}}(\psi)$ upto coupling constant renormalization. 
The scaling dimension of $u$ is estimated numerically and shown to be negative. 
Thus, the Ising potential is irrelevant at $T_c$, and the universality class and the critical exponents in Table \ref{T2} are the same in both cases with the Ising and XY symmetries. 
The universality class of the critical point is dubbed deconfined Ising/XY (DC-Ising/XY) class. \\

\begin{table}[t]
\renewcommand{\arraystretch}{1.2}
\begin{tabular} {c|cccccc}
\hline 
\hline
Universality class& $\alpha$  & ~~$\beta$~~  & ~~$\gamma$~~ & ~~$\nu$~~ & ~~$\eta$~~ & ~~$\delta$~~  \\ \hline 
WF - Ising & $0.11$ & $0.33$ & $1.24$ & $0.63$ & $0.036$ & $4.79$  \\ 
WF - XY  & $-0.015$ & $0.35$ & $1.32$ & $0.67$ & $0.038$ & $4.78$ \\
 \hline
DC - Ising / XY  & $-0.015$    & $0.83$ & $0.35$ & $0.67$ & $1.47$& $1.43$  \\ \hline \hline
\end{tabular}
\caption{\textbf{Critical exponents of the thermal universality classes associated with Ising and XY symmetries in three spatial dimensions.} 
Using the two critical exponents, $\nu$ and $\beta$, we find the other exponents with the scaling relations, $\nu d = 2- \alpha$, $\alpha+2\beta+\gamma=2$, $\gamma = \nu(2-\eta)$, and $\beta(\delta+1)=\nu d$ with $d=3$.
}\label{T2}
\end{table}

{\it Discussion and Conclusion : }
Our analysis with the exactly solvable model of $\hat{H}_{X}$ allows us to include additional microscopic quantum interactions such as $\frac{\textcolor{black}{V_{\theta}}}{2} \sum_j \hat{n}_j^2$ or $h_z \sum_l \hat{\sigma}_l^z$. Note that such inclusions are subtle in the conventional effective thermal gauge theories in the sense that precise couplings are not directly determined. 
Performing perturbative calculations with the other interactions, we check that that the charging effect term with $\textcolor{black}{V_{\theta}}$ induces quantum fluctuations of the rotors, similar to the conventional Bose-Hubbard model. Also, the local gauge invariance breaking term with $h_z$ is shown to be irrelevant for $h_z \ll J$ (See \textcolor{cyan}{SM} \cite{SM}). 
Thus, we argue that the exotic thermal transitions are intact under the other interactions which break the exact solvability and local $\mathbb{Z}_2$ transformation.  

Based on our analysis,  we provide a schematic phase diagram in Fig. \ref{F1}\textcolor{red}{(b)}, expecting that the phase diagram is asymptotically exact if $\Lambda_c^0$ is much bigger than the other energy scales. We also compare our results with previous literature with effective thermal gauge theories in \textcolor{cyan}{SM} \cite{SM} , especially with the seminal works by  Lammert, Rokhsar, and Toner \cite{heisenberg1,heisenberg2}.    
Generalization to different symmetry groups is straightforward. 
One important case is the two Ising symmetries observed in the recent experiments on two nematicity in strongly correlated systems such as doped RbFe$_2$As$_2$ \cite{Rb3,Rb4}. 
Recently, the origin of the B$_{2g}$ nematic order has been considered in several theoretical works \cite{Rb_theory1,Rb_theory2}, but there are still puzzlings in the experiments such as the absence of specific heat anomaly. 
To our best knowledge, the puzzling has not been resolved yet.  
We provide one alternative scenario that the B$_{2g}$ nematicity in doped RbFe$_2$As$_2$ is described by the DC-Ising class.

The scenario makes the following predictions.  
First, there is an additional thermal transition at $T = O(\Lambda_c)$ whose scale is much higher than the onset temperature of the B$_{2g}$  nematicity.
The energy scale of $\Lambda_c$ is expected to be a function of the microscopic interactions such as the Coulomb interaction and Hund coupling. 
Specific heat experiments at higher temperatures would be useful.
Second, the singularity of specific heat at the onset temperature of the B$_{2g}$ nematicity is much milder than the one of the Ising class, as manifested in $\alpha_{\mathrm{DC-Ising\!\!}}=-0.015$. 
The presence of other degrees of freedom such as phonons would make the anomaly of the specific heat invisible in experiments.  
We believe that the milder singularity is one plausible explanation of the absence of discernible specific heat jump in the  the recent experiment on doped RbFe$_2$As$_2$. 
In \textcolor{cyan}{SM} \cite{SM}, the absence of the jump is shown to be impossible in the WF classes, even with fermionic excitations. 
Third, the negative value of $\alpha_{\mathrm{DC-Ising}}$ further indicates that the DC-Ising class is much more stable under disorder and acoustic phonon couplings than the Ising class based on the Harris criterion \cite{Harris} and the Larkin-pikin criterion \cite{LP1}. 
Fourth, the exponent of the order parameter onset (susceptibility), $\beta_{\mathrm{DC-Ising\!\!}}=0.83$ ($\gamma_{\mathrm{DC-Ising\!\!}}=0.35$), is much larger (smaller) than the mean-field and WF universality classes which may be tested in experiments.

In conclusion, we show the existence of exotic thermal transitions with spontaneous symmetry breaking from topological orders.  
All critical exponents of the exotic universality classes are evaluated, and differences from the conventional mean-field and WF classes are emphasized. 
We provide smoking-gun experiments to test the exotic thermal transitions in plausible connections with doped RbFe$_2$As$_2$. \\

{\it Acknowledgement }:
The authors thank H. Kontani and T. Shibauchi for invaluable discussions and comments.  
This work is supported by the National Research Foundation of Korea (NRF) grant No. 2019M3E4A1080411, No.2020R1A4A3079707, and No.2021R1A2C4001847.

\bibliographystyle{apsrev4-1}
%
\onecolumngrid
\clearpage

\setcounter{equation}{0}
\setcounter{figure}{0}
\setcounter{table}{0}
\setcounter{page}{1}

\maketitle 
\makeatletter
\renewcommand{\theequation}{S\arabic{equation}}
\renewcommand{\thefigure}{S\arabic{figure}}
\renewcommand{\thetable}{S\arabic{table}}

\begin{center}
\vspace{10pt}
\textbf{\large Supplemental Material for \\``Exotic Thermal Transitions with Spontaneous Symmetry Breaking''}
\end{center} 
\begin{center} 
{Hanbit Oh and Eun-Gook Moon$^{\ \textcolor{red}{*}}$}\\
\emph{Department of Physics, Korea Advanced Institute of Science and Technology, Daejeon 305-701, Korea}
\vspace{5pt}
\end{center}
\twocolumngrid

\section{Effects of the coupling with Fermions} \label{ss1}
This section provides detailed information on how the WF classes are affected by Fermi-surfaces,  in connection with the unconventional nematicity in Ba$_{1-x}$Rb$_{x}$Fe$_2$As$_2$. 
Our strategy to investigate the singularity structure of the effective action consists of three steps.
$(i)$ Consider free fermions with Fermi-surfaces and WF classes of two Ising nematicity of (B$_{1g}$,B$_{2g}$) representation.
$(ii)$ Determine the coupling between the fermions and order parameters based on a group theory. 
$(iii)$ Integrate out fermions and find the effective action.

$(i)$ We consider a minimal two-band model in the 2d square lattice of Fe atoms with two orbital degrees of freedom for simplicity \cite{S_Raghu}.
Note that the generalization to 3d lattice is straightforward and the main results do not change due to the presence of the Fermi-surface. 
Introducing a two component spinor $\psi_{\vec{k}}=(c_{\vec{k}, xz},c_{\vec{k},yz})^{T}$ with $d_{xz},d_{yz}$ orbitals, the tight-binding Hamiltonian is
	\begin{eqnarray*}
 H_{0}	&=&\sum_{\vec{k}}\psi_{\vec{k}}^{\dagger} \left[(\epsilon_{+}(\vec{k})-\mu)+\epsilon_{-}(\vec{k})\hat{\sigma}^{z}+\epsilon_{xy}(\vec{k})\hat{\sigma}^{x}\right]\psi_{\vec{k}},
\label{e1}	\end{eqnarray*}
with
\begin{eqnarray*}
&&\epsilon_{x}(\vec{k})=-2t_{1}\cos k_{x}-2t_2 \cos k_{y}-4t_{3}\cos k_x k_y,\\
&&\epsilon_{y}(\vec{k})=-2t_{2}\cos k_{x}-2t_1 \cos k_{y}-4t_{3}\cos k_x k_y,
\end{eqnarray*}
\vspace{-20pt}
\begin{eqnarray*}
&& \epsilon_{\pm}(\vec{k})=\frac{\epsilon_{x}(\vec{k})\pm \epsilon_{y}(\vec{k})}{2},\ \epsilon_{xy}(\vec{k})=-4t_{4}\sin k_{x}\sin k_y.
\end{eqnarray*}
The crystal and band structures are illustrated in Fig. \ref{SF1}\textcolor{red}{(a,b)}, with energy eigenvalues, $E_{\pm}(\vec{k})=\epsilon_{+}(\vec{k})-\mu\pm (\epsilon_{xy}(\vec{k})^{2}+ \epsilon_{-}(\vec{k})^{2})^{1/2}$.
Its point group symmetry is D$_{4h}$, and two nematic orders $(\Phi_{1},\Phi_{2})$ in (B$_{1g}$,B$_{2g}$) representations are considered with the potentials, 
\begin{eqnarray}
V_{I}=-\Big[\sum_{j}u_{1}\Phi_{1,j}^2+u_{2}\Phi_{2,j}^2\Big]. \nonumber
\end{eqnarray}

$(ii)$ The coupling between fermions and nematic orders is described by the Yukawa coupling,  
\begin{eqnarray*}
H_{Y}&=&\sum_{\vec{k}}\psi_{\vec{k}}^{\dagger}
\Big[
g_{1}\Phi_{1}\hat{\Gamma}_{1}(\vec{k})+g_{2}\Phi_{2}\hat{\Gamma}_{2}(\vec{k})
\Big]
\psi_{\vec{k}},
\end{eqnarray*} 
and the vertex terms are determined by the point group symmetry, 
 \begin{eqnarray*}
\hat{\Gamma}_{1}(\vec{k})&=&a_{1}(\cos k_{x}-\cos k_{y})+b_{1}\hat{\sigma}^{z},\\
\hat{\Gamma}_{2}(\vec{k})&=&a_{2}( \sin k_{x}\sin  k_{y})+b_{2}\hat{\sigma}^{x} . \label{e2-2} 
 \end{eqnarray*}
The phenomenological constants ($a_{i}, b_{i}$) are introduced.\\ 

$(iii)$ Integrating out the fermions by the standard many-body calculations gives the modification of the coupling constants,
\begin{eqnarray}
u_{i} \rightarrow u_{i}+ 
 \Pi_{i}
\big( 
\{t\},\mu, T
\big),
\nonumber
\end{eqnarray}
with a nematic susceptibility,
\begin{eqnarray}
 \Pi_{i}
\big( 
\{t\},\mu, T
\big)= -g_{i}^2T\sum_{\vec{k},k_{n}}\mathrm{Tr}\big[( G_{0}(\vec{k},k_n) \hat{\Gamma}_{i})^{2}
\big],
\end{eqnarray}
where the fermionic Green's function $G_{0}(\vec{k},k_n)$ is used. 
We evaluate the values of $(\Pi_{1},\Pi_{2})$ and their results at $T=10^{-2}|t_{1}|$ are presented in Fig. \ref{SF1}\textcolor{red}{(c)}.

\begin{center}
\vspace{-5pt}
\begin{figure}[t]
\includegraphics[scale=1.6]{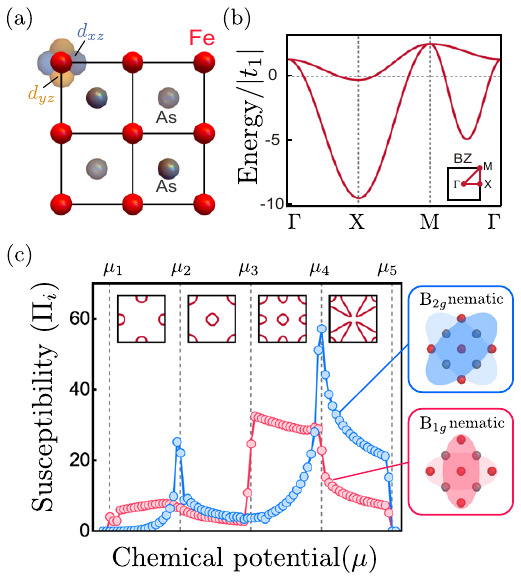}
\caption{
\textbf{(a) Crystal structure of FeAs in the two-dimensional square lattice. }
\textbf{(b) Band structure of the tight-binding Hamiltonian.}
We choose the hopping parameters, $(t_{1},t_{2},t_{3},t_{4})=(-1,1.3,-0.85,-0.85)$ for numerical evaluations.
\textbf{(c) Chemical potential dependence of the nematic susceptibilities.}
Here, the unit $g^{2}_{i}/|t_{1}|$ is adopted.
The numerical values are obtained at $T=10^{-2}|t_{1}|$ and $a_{1}=a_{2}=0,b_{1}=b_{2}=1$.
The Lifhshitz transition appears at $(\mu_{1},\mu_{2},\mu_{3},\mu_{4},\mu_{5})=(-9.45,-4.85,-0.25,1.35,2.55)$
The corresponding electron/hole pocket at each range of chemical potential are illustrated on the top panel.
 } \label{SF1}
\end{figure}
\end{center}
We remark that no additional singularities in the two nematic susceptibilities appear after integrating out Fermi-surfaces for a fixed chemical potential. 
Thus, the universality classes associated with the two nematicities are not different from the conventional WF-Ising classes, and the absence of the specific heat anomaly cannot be explained by coupling to Fermi-surfaces. 
We further note that the relative strength of the two susceptibilities depends on chemical potential. 
It is because the nematic susceptibilities are determined by the shape of Fermi-surfaces, which indicates that doping is an essential factor in determining the properties of the nematicity in this system.

\section{Details on $\hat{H}_{X}$}
This section provides the detailed information on the model Hamiltonian, $\hat{H}_{X}$,
\begin{eqnarray}
\hat{H}_X =  -  J \sum_{\langle i, j \rangle} \hat{\sigma}_{ij}^z \cos( \frac{\hat{\theta}_i}{2} -\frac{\hat{\theta}_j}{2}) -\sum_j  e^{i 2\pi \hat{n}_j} \hat{\mathcal{A}}_j - \sum_{p^{*}} \hat{\mathcal{B}}_{p^*} .\nonumber 
\end{eqnarray}
with the star and plaquette operators, $ \hat{\mathcal{A}}_j = \prod_{l \in j} \hat{\sigma}_l^x$, $\hat{\mathcal{B}}_{p^*} = \prod_{l \in p^*} \hat{\sigma}_l^z$ of the 3d toric code. 
We show the excitation energy of $\hat{H}_{X}$ and the perturbative calculation for checking the stability of the exotic thermal transition. 
\begin{center}
\begin{figure}[t]
  \includegraphics[scale=1.6]{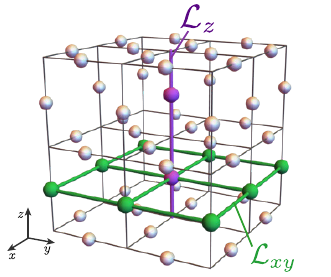}
\caption{\textbf{Two examples of non-local operators, $\hat{\mathcal{V}}_{xy}=\prod_{l\in \mathcal{L}_{xy}}\hat{\sigma}^{x}_{l}$, $\hat{\mathcal{W}}_{z}=\prod_{l\in \mathcal{L}_{z}}\hat{\sigma}^{z}_{l}$,  
distinguishing homologically different ground-states of the pure 3d toric code.}
The presence of the coupling $J$ significantly changes the ground state structures.
The ground state degeneracy is lifted and the state, $\hat{\mathcal{V}}_{xy}|\tilde{\theta}_{0}\rangle$, is no longer a ground state of our model $\hat{H}_{X}$.
\label{SF2}
}
\end{figure}
\end{center}
\subsection{Excitations and energetics of $\hat{H}_{X}$ }
\begin{center}
\begin{figure}[t]
 \includegraphics[scale=1.5]{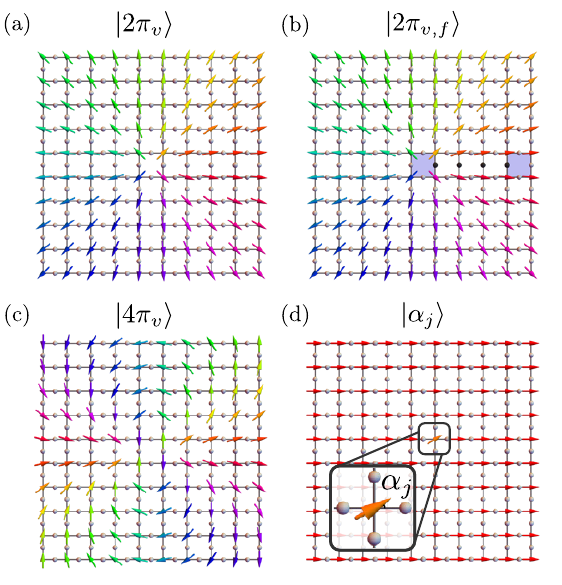} \vspace{-15pt}
\caption{\textbf{
Four possible excitations of $\hat{H}_{X}$.}
The white (black) spheres placed on the link $l$ is a qubit with a quantum number, $\sigma^{z}_{l}=1(-1)$. 
The arrow at the site $j$ is a quantum rotor with an angle variable, $\theta_j$. 
Only the 4$\pi$ vortex-line state and the phase fluctuation 
would be essential to describe the low energy physics of $\hat{H}_{X}$. 
 }\label{SF3}
\end{figure}
\end{center}

We first consider (2$\pi$, 4$\pi$) vortex-line states whose cores are at the center $\mathcal{O}$ of 2d layer, 
\begin{eqnarray*}
|2\pi_{v}\rangle = e^{i\sum_{j}\theta_{v,j}\hat{n}_{j}}|{\tilde \theta}_0\rangle,
\quad
|4\pi_{v}\rangle = e^{i2\sum_{j}\theta_{v,j}\hat{n}_{j}}|{\tilde \theta}_0\rangle,
\end{eqnarray*}
with $\theta_{v,j}=\arctan\left(\frac{y_j}{x_j} \right)$ illustrated in Fig.\;\ref{SF3}\textcolor{red}{(a,c)}.
The excitation energies of the vortex-lines are estimated as,  
\begin{eqnarray*}
\Delta E_{2\pi_{v}}&=&
-J\sum_{
\langle i,j\rangle 
}\Big[
\cos( \frac{\theta_{v,i}}{2} -\frac{\theta_{v,j}}{2}) -1\Big],\\
\Delta E_{4\pi_{v}}&=&
-J\sum_{
\langle i,j\rangle 
}\Big[
\cos( \theta_{v,i} -\theta_{v,j}) -1\Big].
\end{eqnarray*}
It is useful to take a continuum limit, $\theta_{v,j}\rightarrow \theta_{v}(\vec{r})$, and describe the low energy physics. 
The energy difference per unit length ($\mathcal{E}\equiv E/(L_x/a)$) is approximated as \cite{S_subir,S_nagaosa}, 
\begin{eqnarray}
\Delta \mathcal{E}_{2\pi_v}=J\Big[
\frac{\pi}{4}\log
 \left( \frac{L_x}{a}\right)+ \frac{L_x}{a}
 \Big],
\end{eqnarray}\vspace{-10pt}
\begin{eqnarray}
\Delta \mathcal{E}_{4\pi_v}=J
\Big[\pi \log
 \left( \frac{L_x}{a}\right)
 \Big],
\end{eqnarray}
where an additional $L_x/a$ factor of the 2$\pi$ vortex is included due to the discontinuity at $(x>\mathcal{O}_{x},y=\mathcal{O}_{y})$ line. 
Counting the number of possible vortex core estimates the entropy contribution.
The free energy difference per unit length ($f\equiv F/(L_x/a)$) is
\begin{eqnarray*}
\Delta f_{2\pi_v}=J\Big[\frac{\pi}{4}\log
 \left( \frac{L_x}{a}\right)+ \frac{L_x}{a} \Big]-2T\log\left(\frac{L_x}{a} \right),
\end{eqnarray*}\vspace{-10pt}
\begin{eqnarray*}
\Delta f_{4\pi_{v}}=J\Big[\pi \log
 \left( \frac{L_x}{a}\right)\Big]-2T\log\left(\frac{L_x}{a} \right),
\end{eqnarray*}
manifesting that a 4$\pi$ vortex-line is energetically favored compared to a 2$\pi$ vortex-line, in the thermodynamic limit.
The insertion of a pair of flux onto a 2$\pi$ vortex line, $|2\pi_{v,f}\rangle= \prod_{l\in \mathrm{flux}}\hat{\sigma}^{x}_{l}|2\pi_{v}\rangle$, makes the energy lower,
\begin{eqnarray}
\Delta \mathcal{E}_{2\pi_v,f}=J\Big[
\frac{\pi}{4}\log
 \left( \frac{L_x}{a}\right)
 \Big]+4.
\end{eqnarray}
However, the energy cost is still bigger than that of a 4$\pi$ vortex-line in the thermodynamic limit. 
To be specific, if the condition
$\frac{3\pi}{16} \log\left( \frac{L_x}{a}\right)<J^{-1}<\frac{L_x}{4a}$ is satisfied, the energy hierarchy $\Delta \mathcal{E}_{4\pi_v} <\Delta \mathcal{E}_{2\pi_v,f}<
\Delta \mathcal{E}_{2\pi_v}$ is established. 
We emphasize that this inequality is hard to violate in the real system. 
The system size is estimated to be infinitely large to violate the inequality, (i.e. $L_x/a \gtrsim 10^{73}$ for the case of $J=10^{-2}$).
Therefore we conclude that considering only the 4$\pi$ vortex-line state 
along with the phase fluctuation (Fig.\ref{SF3}\textcolor{red}{(d)}) is enough to
describe a low energy Hilbert space of $\hat{H}_{X}$.

\begin{table}[t]
\renewcommand{\arraystretch}{1.45}
\begin{tabular}{C{0.4\linewidth}C{0.55\linewidth}}
\hline 
\hline
Excitation & Wave function    \\ \hline
A pair of fluxes
   & $|\sigma_f \rangle\equiv \prod_{l\in \mathrm{flux}}\sigma^{x}_{l}| {\tilde \theta}_0  \rangle$     \\ 
$2\pi\;$vortex  & $|2\pi_{v}\rangle\equiv  \exp(i\sum_{j}\theta_{v,j}\hat{n}_{j})|{\tilde \theta}_0\rangle$    \\
$2\pi\;$vortex+ fluxes    &$|2\pi_{v,f}\rangle\equiv \prod_{l\in \mathrm{flux}}\hat{\sigma}^{x}_{l}|2\pi_{v}\rangle$\\ 
$4\pi\;$vortex   & $|4\pi_{v}\rangle\equiv  \exp(2i\sum_{j}\theta_{v,j}\hat{n}_{j})|{\tilde \theta}_0\rangle$   \\
Phase fluctuation & $|\alpha_{j}\rangle\equiv
\exp(i\alpha_{j}\hat{n}_{j})|\tilde{\theta}_{0}\rangle$
\\ \hline \hline
\end{tabular}
\caption{\textbf{Wave functions of excited states of $\hat{H}_X$. }
} \label{ST1}
\end{table}

\subsection{Stability under the  perturbation}
We consider gauge symmetry breaking perturbation, $\hat{H}_{\lambda}=\sum_{l}\hat{\sigma}^{z}_{l}$ to check the stability of the exotic thermal transition. 
Under the perturbation, both ground state ($|G\rangle\equiv 2^{\frac{L^3-1}{2}} |\tilde{\theta}_0\rangle$) and ground state energy ($E_{G}$) are modified,
\begin{eqnarray*}
|G\rangle\rightarrow |G \rangle +
h_z
\sum_{ \beta\neq G}
\frac{\langle\beta| \hat{ H}_{\lambda}| G\rangle}{E_{G}-E_{\beta
}}
|\beta\rangle +O(h_z^2),
\end{eqnarray*}\vspace{-5pt}
\begin{eqnarray*}
E_G
&\rightarrow &
E_{G}
+h_z \langle G| \hat{ H}_{\lambda}|  G\rangle+
h_z^2 \sum_{ \beta\neq G}
\frac{|\langle G | \hat{ H}_{\lambda}| \beta \rangle|^2 }{E_{G}-E_{\beta}}
+O(h_z^3), 
\end{eqnarray*}
where the higher order terms of $h_{z}$ are omitted in $O(h_{z}^{2})$, $O(h_{z}^{3})$.
Here, $(|G\rangle,|\beta\rangle )$ are normalized ground/excited states, respectively. 
We find that the leading order correction becomes, 
\begin{eqnarray}
|G\rangle 
&\rightarrow&
|G\rangle 
-\frac{h_z}{4}
\sum_{l}\hat{\sigma}^{z}_{l}
|G\rangle 
+O(h_z^2),\\
E_{G}&\rightarrow & 
E_{G}-\frac{h_z^2}{4}
\times
(3L^3)
+O(h_z^3).
\end{eqnarray}

Our perturbative analysis demonstrates that the ground state of $\hat{H}_{X}$ is stable and our schematic phase diagram (Fig.\;\ref{F1}\textcolor{red}{(b)} in the main-text) does not affected if $ h_z $ is smaller than other energy scales.  
\begin{center}
\begin{figure}[t]
 \includegraphics[scale=1.45]{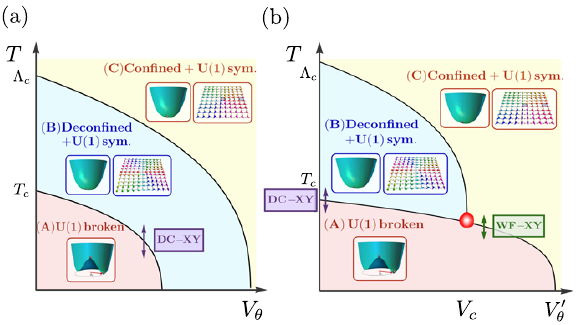}\quad
\caption{ \textbf{(a),(b) Two schematic phase diagrams associated with deconfined phases.} 
Two symmetric phases (B,C) are distinguished by different topological defects. 
While a $4\pi$ vortex line is a fundamental excitation in the deconfined phase, a $2\pi$ vortex line is also allowed in the confined phase.
The two transitions associated with a symmetry breaking and deconfinment $(\Lambda_{c}, T_{c})$ are illustrated in each diagram. 
}
\label{SF4}
\end{figure}
\end{center}

\subsection{Scenarios of symmetry breaking transitions
associated with deconfined phases}
In the main-text, we illustrate the case where the transition associated with the symmetry-breaking and deconfinement are well separated,
 $\Lambda_{c}\gg T_{c}$ (See Fig. \ref{SF3}\textcolor{red}{(a)}).
However, another alternative scenario is possible, in which the order of energy scale associated with two transitions is reversed (See Fig. \ref{SF3}\textcolor{red}{(b)}).
The critical behaviors of the two transitions at $\textcolor{black}{V_{\theta}'}<V_{c}$ and $\textcolor{black}{V_{\theta}'}>V_{c}$, are qualitatively different.
\begin{table}[t]
\renewcommand{\arraystretch}{1.2}
\begin{tabular} {c|c c c c c c}
\hline \hline
Universality class& $\alpha$  & ~~$\beta$~~  & ~~$\gamma$~~ & ~~$\nu$~~ & ~~$\eta$~~ & ~~$\delta$~~  \\ \hline 
WF - Heisenberg & $-0.12$ & $0.37$ &  $1.40$ & $0.71$ & $0.035$& $1.54$  \\ \hline  
DC - Heisenberg  & $-0.12$    & $0.84$ & $0.44$ & $0.71$ & $1.38$& $0.035$  \\ \hline \hline
\end{tabular}
\caption{\textbf{Critical exponents of the two thermal universality classes associated with the $O(3)$ symmetry in three spatial dimensions. }
Using the two critical exponents, $\nu$ and $\beta$, we find the other exponents with the scaling relations, $\nu d = 2- \alpha$, $\alpha+2\beta+\gamma=2$, $\gamma = \nu(2-\eta)$, and $\beta(\delta+1)=\nu d$ with $d=3$.
}\label{ST2}
\end{table}

\section{Comparison with previous works}
Our approach to the exotic universality classes has a similarity to the previous literature \cite{S_heisenberg1,S_heisenberg2}, where a nematic transition associated with the Ising gauge structure was considered.
Via Monte Carlo calculations, they concluded that the universality class is of the Heisenberg spin model.
The similarity is originated from the fact that an order parameter is not a primary operator but a secondary one. 
Still, we stress that our approach does not employ any gauge invariance but provides concrete microscopic degrees of freedom. 
Thus, the origin of the gauge invariance is not an issue in our approach. 
We consider the universality class of the $O(3)$ rotors coupled to the toric code,  
\begin{eqnarray}
H_{O(3)} = - J \sum_{ \langle i,j \rangle}\hat{ \sigma}_{ij}^z \hat{s}_i^{a} \hat{s}_j^a -\sum_j \hat{ U}_{j}\hat{\mathcal{A}}_j - \sum_{p^{*}}\hat{ \mathcal{B}}_{p^*}, 
\end{eqnarray}
where the inversion operator of a rotor at a site $j$ is introduced, $\hat{U}_j$, whose action gives
\begin{eqnarray}
\hat{U}_j^{\dagger} \hat{s}_j^a \hat{U}_j = -\hat{s}_j^a. \nonumber
\end{eqnarray}
In Table \ref{ST2}, all the critical exponents are listed. 
Note that our result, $\beta_{\mathrm{DC-Heisenberg}\!\!}\sim 0.84$ is different from the Heisenberg one, $\beta_{\mathrm{WF-Heisenberg}\!\!}\sim 0.37$.

\bibliographystyle{apsrev4-1}

\end{document}